\documentclass[aps,prb,superscriptaddress,twocolumn,showkeys]{revtex4-1}
\usepackage{color}
\usepackage{graphicx}
\usepackage{amsmath}
\usepackage{amssymb}
\usepackage{url}
\usepackage{soul}
\begin{document}

\title{Multi-Valley Superconductivity In Ion-Gated MoS$_2$ Layers}

\author{Erik Piatti}
\thanks{These authors contributed equally to this work.}
\affiliation{Department of Applied Science and Technology, Politecnico di Torino, 10129 Torino, Italy}
\author{Domenico De Fazio}
\thanks{These authors contributed equally to this work.}
\affiliation{Cambridge Graphene Centre, University of Cambridge, Cambridge CB3 OFA, UK}
\author{Dario Daghero}
\affiliation{Department of Applied Science and Technology, Politecnico di Torino, 10129 Torino, Italy}
\author{Srinivasa Reddy Tamalampudi}
\author{Duhee Yoon}
\author{Andrea C. Ferrari}
\affiliation{Cambridge Graphene Centre, University of Cambridge, Cambridge CB3 OFA, UK}
\email{acf26@eng.cam.ac.uk}
\author{Renato S. Gonnelli}
\affiliation{Department of Applied Science and Technology, Politecnico di Torino, 10129 Torino, Italy}


\begin{abstract}
Layers of transition metal dichalcogenides (TMDs) combine the enhanced effects of correlations associated with the two-dimensional limit with electrostatic control over their phase transitions by means of an electric field. Several semiconducting TMDs, such as MoS$_2$, develop superconductivity (SC) at their surface when doped with an electrostatic field, but the mechanism is still debated. It is often assumed that Cooper pairs reside only in the two electron pockets at the K/K' points of the Brillouin Zone. However, experimental and theoretical results suggest that a multi-valley Fermi surface (FS) is associated with the SC state, involving 6 electron pockets at the Q/Q' points. Here, we perform low-temperature transport measurements in ion-gated MoS$_2$ flakes. We show that a fully multi-valley FS is associated with the SC onset. The Q/Q' valleys fill for doping$\gtrsim2\cdot10^{13}$cm$^{-2}$, and the SC transition does not appear until the Fermi level crosses both spin-orbit split sub-bands Q$_1$ and Q$_2$. The SC state is associated with the FS connectivity and promoted by a Lifshitz transition due to the simultaneous population of multiple electron pockets. This FS topology will serve as a guideline in the quest for new superconductors.
\end{abstract}

\maketitle

Transition metal dichalcogenides (TMDs) are layered materials with a range of electronic properties. Depending on chemical composition, crystalline structure, number of layers (N), doping, and strain, different TMDs can be semiconducting, metallic and superconducting\cite{FerrN2015}. Amongst semiconducting TMDs, MoS$_2$, MoSe$_2$, WS$_2$ and WSe$_2$ have sizeable bandgaps in the range$\sim$1-2eV\cite{WangNN2012}. When exfoliated from bulk to single layer (1L), they undergo an indirect-to-direct gap transition\cite{MakPRL2010,SpleNL2010,WangNN2012}, offering a platform for electronic and optoelectronic applications\cite{FerrN2015,WangNN2012,MakNP2016}, such as transistors\cite{PodzAPL2004,RadiNN2011,FangNL2012}, photodetectors\cite{GourSEMSC1997, LeeNL2012, YinACS2012, KoppNN2014}, modulators\cite{SunNP2016} and electroluminescent devices\cite{CarlPRB2002,SundNL2013}.

\begin{figure*}
\centerline{\includegraphics[width=0.8\textwidth]{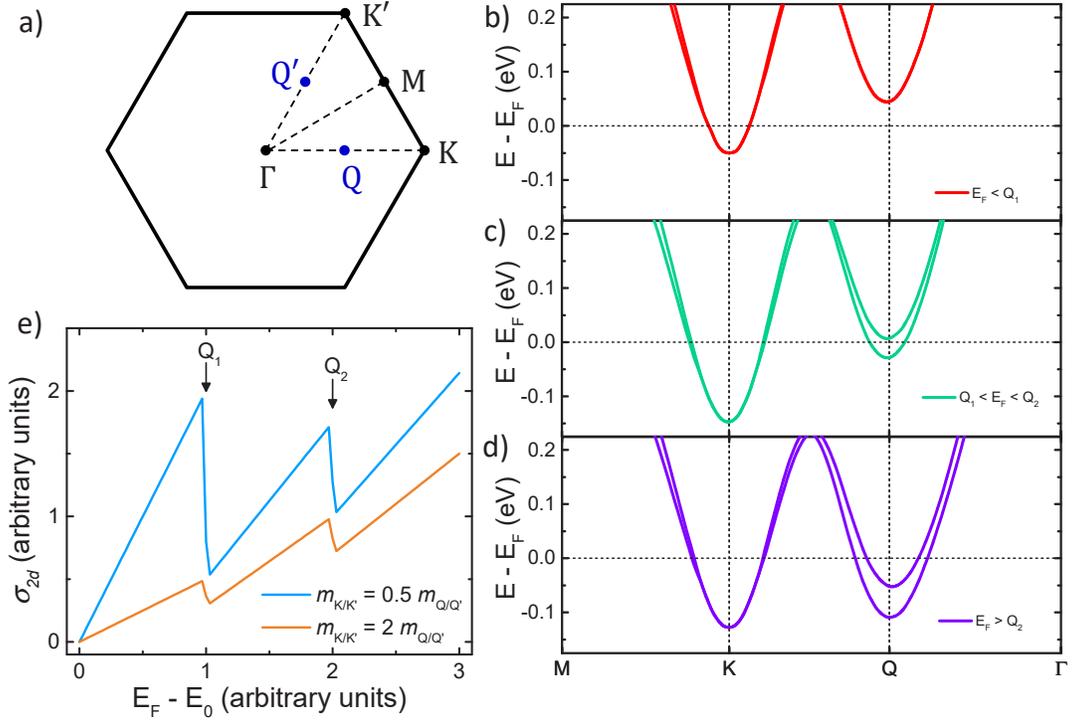}}
\caption{a) BZ for the 2H crystal structure. High symmetry points $\Gamma$, M, K, K', and points Q and Q' are indicated. b-d) 3L-MoS$_2$ band structure for 3 doping values. The bands are adapted from Ref.\citenum{BrummePRB2015}, and were obtained by Density Functional Theory (DFT) calculations using the Quantum ESPRESSO package\cite{GiannozziJPCM2017} in the field-effect transistor configuration\cite{BrummePRB2014}. The valleys at K/K' have two spin-orbit split sub-bands, with splitting much smaller than at Q/Q', not seen in this scale. e) Schematic dependence of $\sigma_{2d}$ for increasing E$_F$ above the global energy minimum of the conduction band, $E_0$, at T=0, for 2 ratios of the effective masses in the K/K' and Q/Q' electron pockets ($m_\mathrm{K/K'}$ and $m_\mathrm{Q/Q'}$). Curves are calculated using Eq.6 of Ref.\citenum{BrummePRB2016} and setting a total degeneracy of 4 for the K/K' pockets, 6 for the first Q/Q' pocket (Q$_1$), and 6 for the second Q/Q' pocket (Q$_2$); physical constants and energy separations are set to unity.}
\label{figure:bands}
\end{figure*}

For all TMDs with 2H crystal structure, the hexagonal Brillouin Zone (BZ) features high-symmetry points $\Gamma$, M, K and K'\cite{SpleNL2010,BrummePRB2015}, Fig.\ref{figure:bands}a. The minima of the conduction band fall at K, K', as well as at Q, Q', approximately half-way along the $\Gamma$-K(K') directions\cite{SpleNL2010,BrummePRB2015}, Fig.\ref{figure:bands}a. In absence of an out-of-plane electric field, the relative position of Q and Q' depends on N and strain\cite{SpleNL2010,LambrechtPRB2012,BrummePRB2015}. The global minimum of the conduction band sits at K/K' in 1L-MoS$_2$ and at Q/Q' in few layer (FL)-MoS$_2$ with N$\geq$4\cite{SpleNL2010}. When an electric field is applied perpendicular to the MoS$_2$ plane, inversion symmetry is broken and the global minimum of the conduction band is shifted to K/K' in any FL-MoS$_2$\cite{BrummePRB2015}, Figs.\ref{figure:bands}b-d. The valleys at K/K' and at Q/Q' are characterized by a different electron-phonon coupling (EPC)\cite{GePRB2013} and, when inversion symmetry is broken, by a different spin-orbit coupling (SOC)\cite{KadantsevSSC2012}. In particular, both EPC and SOC are larger in the Q/Q' valleys\cite{GePRB2013,KadantsevSSC2012}.

The field-effect transistor (FET) architecture is ideally suited to control the electronic properties of 1L flakes, as it simultaneously provides an electrostatic control of the transverse electric field and the carrier density. In the electric-double-layer (EDL) technique\cite{FujimotoReview2013}, the standard solid gate dielectric is replaced by an ionic medium, such as an ionic liquid or electrolyte. In this configuration, the EDL that forms at the ionic liquid/electrode interfaces supports electric fields in excess of$\sim$10MV/cm\cite{UenoReview2014}, corresponding to surface carrier densities $n_{2d}\gtrsim10^{14}$cm$^{-2}$\cite{UenoReview2014}. Ionic-liquid gating has been used to tune the Fermi level, $E_F$, in TMDs and explore transport at different carrier concentrations\cite{SaitoReview2016, YeScience2012, BragNL2012, YuNN2015, XiPRL2016}. The vibrational properties of TMDs can also be controlled by means of the EDL technique, as suggested by gate-induced softening of Raman-active modes in 1L-MoS$_2$\cite{ChakrabortyPRB2012}, while the opposite is observed in gated 1L\cite{DasNatureNanotech2008} and two-layer (2L)\cite{DasPRB2009} graphene. Ref.\citenum{YeScience2012} reported a gate-induced superconducting state at the surface of liquid-gated MoS$_2$ flakes with N$\gtrsim$25\cite{YeScience2012}, while Ref.\citenum{CostanzoNatNano2016} detected this down to N=1.

Most of these results have been interpreted in terms of the population of the conduction band minima at K/K'\cite{YeScience2012, LuScience2015, SaitoNatPhys2016, ChenPRL2017}, which are global minima in both 1L-MoS$_2$\cite{MakPRL2010, GePRB2013} and electrostatically-doped FL-MoS$_2$\cite{YeScience2012, LuScience2015, SaitoNatPhys2016, BrummePRB2015}, Fig.\ref{figure:bands}b. Theoretical investigations however suggested that the population of the high-energy minima at Q/Q' may have an important role in determining the properties of gated MoS$_2$ flakes, by providing contributions both to EPC\cite{GePRB2013, BrummePRB2015} and SOC\cite{KadantsevSSC2012, WuNatCommun2016}. Ref.\citenum{GePRB2013} predicted that when the Q/Q' valleys of 1L-MoS$_2$ are populated (Fig.\ref{figure:bands}c,d), EPC strongly increases (from$\sim0.1$ to$\sim18$), leading to a superconducting transition temperature $T_c\sim20$K for a doping level $x=0.18$ electrons(e$^-$)/unit cell (corresponding to $E_F=0.18\pm0.02$eV at K/K' and $0.08\pm0.02$eV at Q/Q')\cite{BrummePRB2015}). However, Ref.\citenum{CostanzoNatNano2016} measured $T_c\sim2$K for $x\sim0.09\div0.17$ e$^-$/unit cell in e$^-$-doped 1L-MoS$_2$. This mismatch may be associated with the contribution of e$^-$--e$^-$ interactions, whose role in the determination of $T_c$ is still under debate\cite{RoldanPRB2013, DasPRB2015}. Overall, the agreement between the model of Ref.\citenum{GePRB2013} and the trend of $T_c$ with e$^-$ doping in Ref.\citenum{YeScience2012} suggests that the mechanism of Ref.\citenum{GePRB2013} for EPC enhancement when the Q/Q' valleys are crossed may also hold for FL-MoS$_2$.

Inversion symmetry can be broken in MoS$_2$ either by going to the 1L limit\cite{GePRB2013}, or by applying a transverse electric field\cite{KormanyosPRB2013, YuanPRL2014}. This leads to a finite SOC\cite{KormanyosPRB2013, YuanPRL2014}, which lifts the spin degeneracy in the conduction band and gives rise to two spin-orbit-split sub-bands in each valley\cite{KormanyosPRB2013, BrummePRB2015}, as shown in Fig.\ref{figure:bands}b-d for FL-MoS$_2$. When the system is field-effect doped, the inversion symmetry breaking increases with increasing transverse electric field \cite{LuScience2015, ChenPRL2017}, due to the fact that induced e$^-$ tend to become more localized within the first layer \cite{LuScience2015, ChenPRL2017, RoldanPRB2013}. Hence, the SOC and the spin-orbit splitting between the bands increase as well, as was calculated in Ref.\citenum{BrummePRB2015}.

When combined to the gate-induced SC state\cite{YuanPRL2014}, this can give rise to interesting physics, such as spin-valley locking of the Cooper pairs\cite{SaitoNatPhys2016} and 2d Ising superconductivity (SC)\cite{LuScience2015} with a non-BCS-like energy gap\cite{CostanzoNatNano2018}, suggested to host topologically non-trivial SC states\cite{RoldanPRB2013, HsuNatComms2017, NakamuraPRB2017}. Refs.\citenum{BrummePRB2015, KadantsevSSC2012} predicted SOC and spin-orbit splitting between sub-bands to be significantly stronger for the Q/Q' valleys than for K/K', thus supporting spin-valley locking at Q/Q' as well\cite{WuNatCommun2016}. A dominant contribution of the Q/Q' valleys in the development of the SC state would be consistent with the high ($\gtrsim50$T) in-plane upper critical field, $H_{c2}^{||}$, observed in ion-gated MoS$_2$ \cite{LuScience2015, SaitoNatPhys2016} and WS$_2$ \cite{LuPNAS2018}. The $H_{c2}^{||}$ enhancement is caused by locking of the spin of the Cooper pairs in the out-of-plane direction in a 2d superconductor in the presence of finite SOC, and is therefore promoted by increasing the SOC. However, $H_{c2}^{||}$ for MoS$_2$ and WS$_2$ is higher than in metallic TMD Ising superconductors (such as NbSe$_2$ and TaS$_2$), where $H_{c2}^{||}\lesssim30$T \cite{delaBarreraNatCommun2018}, despite the SOC in the K/K' valleys being much smaller ($\sim3$meV for MoS$_2$\cite{KormanyosPRB2013}). Spin-valley locking in the Q/Q' valleys may thus explain this apparent inconsistency in the physics of ion-gated semiconducting TMDs under magnetic field.

From the experimental point of view, the possible multi-valley character of transport in gated TMDs is currently debated. Refs.\citenum{WuNatCommun2016, CuiNatNano2015, ChenPRL2017} measured the Landau-level degeneracy at moderate $n_{2d}\sim10^{12}-10^{13}$cm$^{-2}$, finding it compatible with a carrier population in the Q/Q' valleys. However, Ref.\citenum{ChenPRL2017} argued that this would be suppressed for larger $n_{2d}\gtrsim10^{13}$cm$^{-2}$, typical of ion-gated devices and mandatory for the emergence of SC) due to stronger confinement within the first layer\cite{ChenPRL2017}. In contrast, angle-resolved photoemission spectroscopy in surface-Rb-doped TMDs\cite{KangNanoLett2017} highlighted the presence of a non-negligible spectral weight at the Q/Q' valleys only for $n_{2d}\gtrsim8\cdot10^{13}$cm$^{-2}$ in the case of MoS$_2$. Thus, which valleys and sub-bands are involved in the gate-induced SC state still demands a satisfactory answer.

Here we report multi-valley transport and SC at the surface of liquid-gated FL-MoS$_2$. We use a dual-gate geometry to tune doping across a wide range of $n_{2d}\sim5\cdot10^{12}-1\cdot10^{14}$cm$^{-2}$, induce SC, and detect characteristic ``kinks'' in the transconductance. These are non-monotonic features that emerge in the $n_{2d}$-dependence of the low-temperature ($T$) conductivity when $E_F$ crosses the high-energy sub-bands\cite{BrummePRB2016}, irrespectively of their specific effective masses, Fig.\ref{figure:bands}e. We show that the population of the Q/Q' valleys is fundamental for the emergence of SC. The crossing of the first sub-band Q$_1$ (Fig.\ref{figure:bands}c) occurs at small $n_{2d}\lesssim2\cdot10^{13}$cm$^{-2}$, implying that multi-valley transport already occurs in the metallic phase over a wide range of $n_{2d}\sim2-6\cdot10^{13}$cm$^{-2}$. We also show that the crossing of the second sub-band Q$_2$ occurs after a finite $T_c$ is observed, while a full population of both spin-orbit-split sub-bands (Fig.\ref{figure:bands}d) in the Q/Q' valleys is required to reach the maximum $T_c$. These results highlight how SC can be enhanced in MoS$_2$ by optimizing the connectivity of its Fermi Surface (FS), i.e. by adding extra FSs in different BZ regions to provide coupling to further phonon branches\cite{PickettBook}. Since the evolution of the band structure of MoS$_2$ with field-effect doping is analogous to that of other semiconducting TMDs\cite{BrummePRB2015, BrummePRB2016, DasPRB2015, WuNatCommun2016, KangNanoLett2017}, a similar mechanism is likely associated with the emergence of SC in TMDs in general. Thus, optimization of the FS connectivity can be a viable strategy in the search of new superconductors.
\begin{figure}
\centerline{\includegraphics[width=0.8\columnwidth]{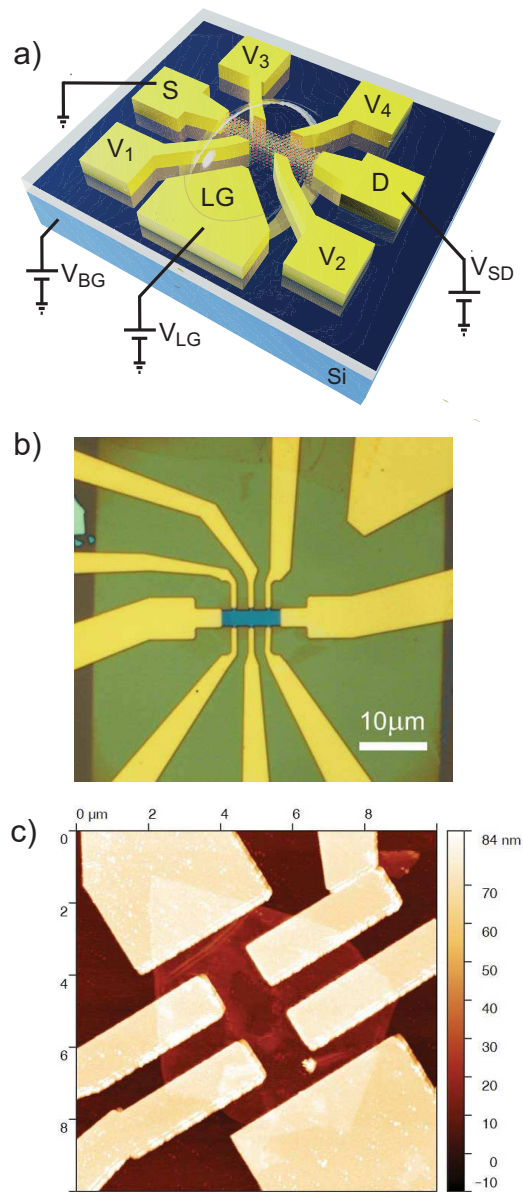}}
\caption{a) Hall bar FL-MoS$_2$ flake with voltage probes ($V_i$), source (S), drain (D) and liquid-gate (LG) electrodes. A ionic liquid droplet covers the flake and part of the LG electrode. The sample is biased with a source-drain voltage ($V_{SD}$) and dual gate control is enabled by a voltage applied on the liquid gate ($V_{LG}$) and on the solid back gate ($V_{BG}$). b) Optical image of Hall bar with six voltage probes. The LG electrode is on the upper-right corner. c) AFM scan of the MoS$_2$ Hall bar after ionic liquid removal.}
\label{figure:device}
\end{figure}

We study flakes with N=4-10, as Refs.\citenum{SpleNL2010, BrummePRB2015, BrummePRB2016} predicted that flakes with N$\geq$4 are representative of the bulk electronic structure, and Ref.\citenum{CostanzoNatNano2016} experimentally observed that both $T_c$ and the critical magnetic field $H_{c2}$ in 4L flakes are similar to those of 6L and bulk flakes. Our devices are thus comparable with those in literature\cite{YeScience2012, LuScience2015, SaitoNatPhys2016, CostanzoNatNano2016, FuQuantMater2017}. We do not consider 1L flakes as they exhibit a lower $T_c$ and their mobility is suppressed due to disorder\cite{CostanzoNatNano2016, FuQuantMater2017}.

FL-MoS$_2$ flakes are prepared by micro-mechanical cleavage\cite{NovoPNAS2005} of 2H-MoS$_2$ crystals from SPI Supplies. The 2H phase is selected to match that in previous reports of gate-induced SC\cite{YeScience2012, CostanzoNatNano2016}. Low resistivity ($<0.005\Omega\cdot$cm) Si coated with a thermal oxide SiO$_2$ is chosen as a substrate. We tested both 90 or 285nm SiO$_2$ obtaining identical SC results. Thus, 90nm SiO$_2$ is used to minimize the back gate voltage $V_{BG}$ \hbox{(-30V$<V_{BG}<$30V)}, while 285nm is used to minimize leakage currents through the back gate $I_{BG}$. Both SiO$_2$ thicknesses provide optical contrast at visible wavelengths\cite{CasiraghiNanoLett2007}. A combination of optical contrast, Raman spectroscopy and atomic force microscopy (AFM) is used to select the flakes and determine N.

Electrodes are then defined by patterning the contacts area by e-beam lithography, followed by Ti:$10$nm/Au:$50$nm evaporation and lift-off. Ti is used as an adhesion layer\cite{Kutz2002}, while the thicker Au layer provides the electrical contact. Flakes with irregular shapes are further patterned in the shape of Hall bars by using polymethyl methacrylate (PMMA) as a mask and removing the unprotected MoS$_2$ with reactive ion etching (RIE) in a $150$mTorr atmosphere of CF$_4$:O$_2$=5:1, as shown in Figs.\ref{figure:device}a,b. A droplet of 1-Butyl-1-methylpiperidinium bis(trifluoromethylsulfonyl)imide (BMPPD-TFSI) is used to cover the FL-MoS$_2$ surface and part of the side electrode for liquid gate operation (LG), as sketched in Fig.\ref{figure:device}a.

AFM analysis is performed with a Bruker Dimension Icon in tapping mode. The scan in Fig.\ref{figure:device}c is done after the low-T experiments and removal of the ionic liquid, and confirms that the FL-MoS$_2$ sample does not show topographic damage after the measurement cycle.
\begin{figure}
\centerline{\includegraphics[width=\columnwidth]{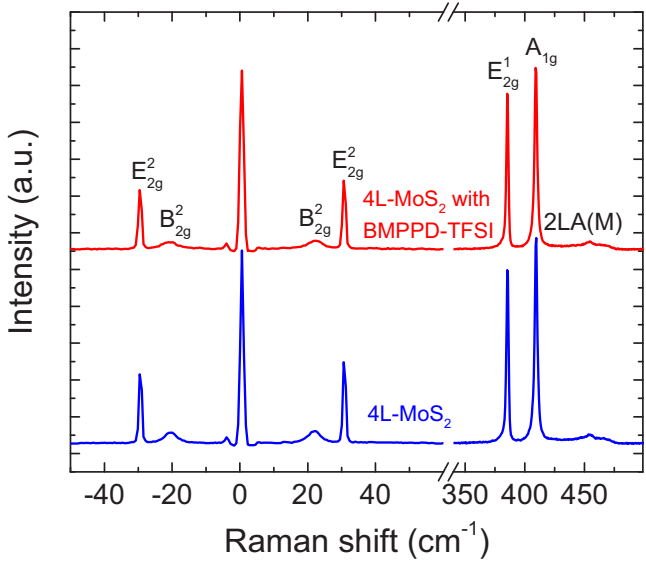}}
\caption{Representative Raman spectra at 514nm of a 4L-MoS$_2$ flake before (blue) and after (red) device fabrication, deposition of the ionic liquid droplet and low-T transport measurements.}
\label{figure:Raman}
\end{figure}

We use Raman spectroscopy to characterize the devices both before and after fabrication and BMPPD-TFSI deposition. Raman measurements are performed with a Horiba LabRAM Evolution at 514nm, with a 1800grooves/mm grating and a spectral resolution$\sim0.45$cm$^{-1}$. The power is kept below 300$\mu$W to avoid any damage. A representative Raman spectrum of 4L-MoS$_2$ is shown in Fig.\ref{figure:Raman} (blue curve). The peak at$\sim$455cm$^{-1}$ is due to a second-order longitudinal acoustic mode at the M point\cite{StacJPCS1985}. The E$_{2g}^1$ peak at$\sim$385cm$^{-1}$ and the A$_{1g}$ at$\sim$409cm$^{-1}$ correspond to in-plane and out-of plane vibrations of Mo and S atoms\cite{VerbPRL1970,WietPRB1971}. Their difference, Pos(E$_{2g}^1$)-Pos(A$_{1g}$), is often used to monitor N\cite{LeeACS2010}. However, for N$\geq$4, the variation in Pos(E$_{2g}^1$)-Pos(A$_{1g}$) between N and N$+$1 approaches the instrument resolution\cite{LeeACS2010} and this method is no longer reliable. Thus, we use the low frequency modes ($<100$cm$^{-1}$) to monitor N\cite{ZhanPRB2013,TanNM2012}. The shear (C) and layer breathing modes (LBM) are due to the relative motions of the atomic planes, either perpendicular or parallel to their normal\cite{ZhanPRB2013}. Pos(C) and Pos(LBM) change with N as\cite{ZhanPRB2013,TanNM2012}:
\begin{equation}
\label{Eq1}
\mathrm{Pos(C)_N}=\frac{1}{\sqrt{2}\pi c}\sqrt{\frac{\alpha_{\parallel}}{\mu_m}}\sqrt{1+\cos\left(\frac{\pi}{N}\right)}
\end{equation}
\begin{equation}
\label{Eq2}
\mathrm{Pos(LBM)_N}=\frac{1}{\sqrt{2}\pi c}\sqrt{\frac{\alpha_{\perp}}{\mu_m}}\sqrt{1-\cos\left(\frac{\pi}{N}\right)}
\end{equation}
where $\alpha_{\parallel}\sim$2.82$\cdot$10$^{19}$N/m$^3$ and $\alpha_{\perp}\sim$8.90$\cdot$10$^{19}$N/m$^3$ are spring constants for C and LBM modes, respectively, $c$ is the speed of light in vacuum, $\mu_m\sim$3$\cdot$10$^{-6}$Kg/m$^2$ is the 1L mass per unit area\cite{ZhanPRB2013,TanNM2012}. Fig.\ref{figure:Raman} shows a C mode at$\sim$30cm$^{-1}$ and an LBM at$\sim$22cm$^{-1}$. These correspond to N$=$4 using Eqs.\ref{Eq1},\ref{Eq2}. Fig.\ref{figure:Raman} also plots the Raman measurements after device fabrication, deposition of the ionic liquid, low-T measurements, $V_{LG}$ removal and warm-up to room T (red curve). We still find Pos(C)$\sim$30cm$^{-1}$ and Pos(LBM)$\sim$22cm$^{-1}$, the same as those of the pristine flake, suggesting no damage nor residual doping.

Four-probe resistance and Hall measurements are then performed in the vacuum chambers of either a Cryomech pulse-tube cryocooler, $T_{min}$=2.7K, or a Lakeshore cryogenic probe-station, $T_{min}$=8K, equipped with a 2T superconducting magnet. A small ($\sim1\mu$A) constant current is applied between S and D (Fig.\ref{figure:device}a) by using a two-channel Agilent B2912A source-measure unit (SMU). The longitudinal and transverse voltage drops are measured with an Agilent 34420 low-noise nanovoltmeter. Thermoelectrical and other offset voltages are eliminated by measuring each resistance value and inverting the source current in each measurement\cite{DagheroPRL2012}. Gate biases are applied between the corresponding G and D with the same two-channel SMU (liquid gate) or a Keithley 2410 SMU (back gate). Samples are allowed to degas in vacuum ($<10^{-5}$mbar) at room $T$ for at least$\sim1$h before measurements, in order to remove residual water traces in the electrolyte.
\begin{figure*}
\centerline{\includegraphics[width=0.8\textwidth]{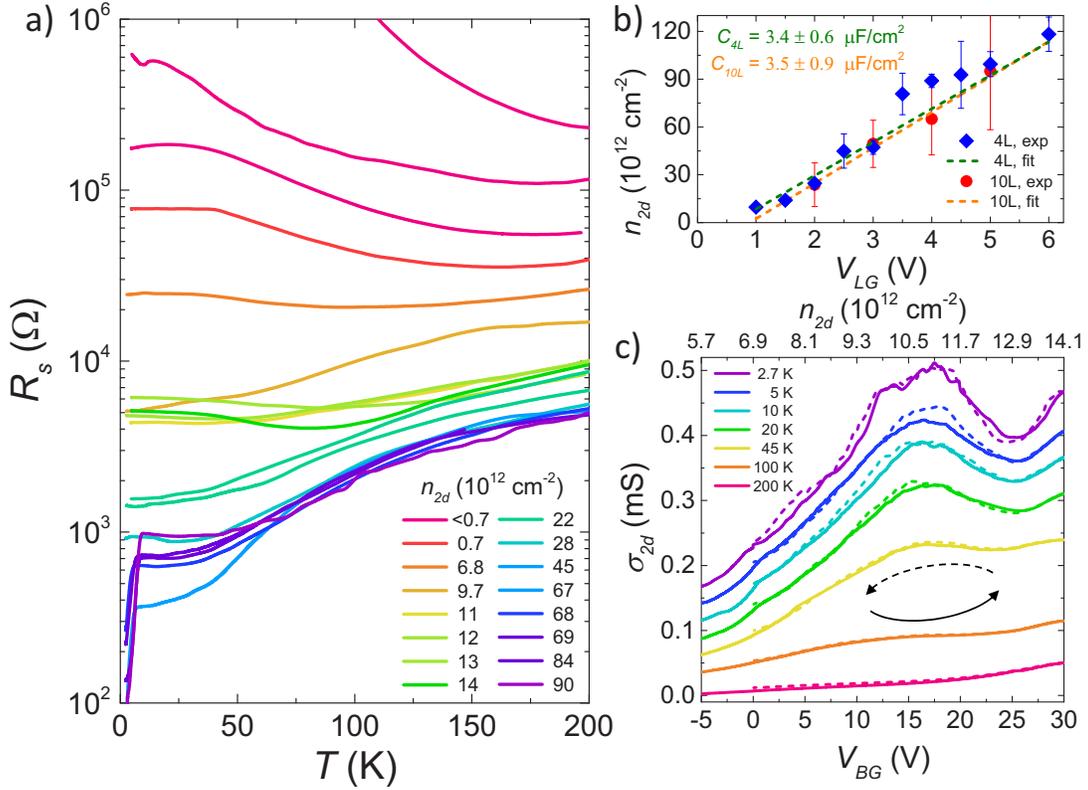}}
\caption{Transport of dual-gated 4L-MoS$_2$. a) $R_s$ as a function of T for different $n_{2d}$. b) $n_{2d}$ as function of $V_{LG}$ as determined via Hall effect measurements, for N=4, 10. The liquid gate capacitances are obtained by a linear fit of the data. c) $\sigma_{2d}$ as a function of $V_{BG}$ at $V_{LG}=0.9$V, for different T. Each curve is shifted by $3.333\times10^{-5}$S. The top scale shows the values of $n_{2d}$ estimated from $C_{ox}$. Solid (dashed) curves are measured for increasing (decreasing) $V_{BG}$.}
\label{figure:transport}
\end{figure*}

We first characterize the T dependence of the sheet resistance, $R_s$, under the effect of the liquid top gate. We apply the liquid gate voltage, $V_{LG}$, at 240K, where the electrolyte is still liquid, and under high-vacuum ($<10^{-5}$mbar) to minimize unwanted electrochemical interactions and extend the stability window of the ionic liquid\cite{UenoReview2014}. After $V_{LG}$ is applied, we allow the ion dynamics to settle for$\sim$10min before cooling to a base T=2.7K.

Fig.\ref{figure:transport}a plots the T dependence of $R_s$ measured in a four-probe configuration, for different $V_{LG}$ and induced carrier density $n_{2d}$. Our devices behave similarly to Ref.\citenum{YeScience2012}, undergoing first an insulator-to-metal transition near $R_s\sim h/e^2$ at low $n_{2d}<1\cdot 10^{13}$cm$^{-2}$, followed by a metal-to-superconductor transition at high $n_{2d}>6\cdot 10^{13}$cm$^{-2}$. The saturating behavior in the $R_s$ vs T curves in Fig.\ref{figure:transport}a for $\mathrm{T}\lesssim50$K, close to the insulator-to-metal transition, is typically observed in systems at low $n_{2d}$ characterized by a fluctuating electrostatic potential, such as that due to charged impurities\cite{ZabrodskiiJETP1984}. This applies to ion-gated crystalline systems at low $V_{LG}$, since the doping is provided by a low density of ions in close proximity to the active channel. These ions induce a perturbation of the local electrostatic potential, locally inducing charge carriers, but are otherwise far apart. The resulting potential landscape is thus inhomogeneous. This low-doping ($\lesssim1\times10^{13}$cm$^{-2}$) density inhomogeneity is a known issue in ion-gated crystalline systems, but becomes less and less relevant at higher ionic densities\cite{RenNanoLett2015}. We employ Hall effect measurements to determine $n_{2d}$ as a function of $V_{LG}$ (see Fig.\ref{figure:transport}b), and, consequently, the liquid gate capacitance $C_{LG}$. $C_{LG}$ for the BMPPD-TFSI/MoS$_2$ interface ($\sim3.4\pm0.6\mu$F/cm$^{2}$) is of the same order of magnitude as for DEME-TFSI/MoS$_2$ in Ref.\citenum{ShiSciRep2015} ($\sim8.6\pm4.1\mu$F/cm$^2$), where DEME-TFSI is the N,N-Diethyl-N-methyl-N-(2-methoxyethyl)ammonium bis(trifluoromethanesulfonyl)imide ionic liquid\cite{ShiSciRep2015}.

Fig.\ref{figure:transport}a shows that, while for T$\gtrsim$100K $R_s$ is a monotonically decreasing function of $n_{2d}$, the same does not hold for T$\lesssim$100K, where the various curves cross. In particular, the residual $R_s$ in the normal state $R_s^0$ (measured just above $T_c$ when the flake is superconducting) varies non-monotonically as a function of $n_{2d}$. This implies the existence of multiple local maxima in the $R_s^0 (n_{2d})$ curve. Consistently with the theoretical predictions of Ref.\citenum{BrummePRB2016}, we find two local maxima. The first and more pronounced occurs when the flake is superconducting, i.e. for $n_{2d}>6\cdot 10^{13}$cm$^{-2}$. This feature was also reported in Refs.\citenum{YeScience2012, LuScience2015}, but not discussed. The second, less pronounced kink, is observed for $1\cdot 10^{13} \lesssim n_{2d} \lesssim 2\cdot 10^{13}$cm$^{-2}$, not previously shown. Both kinks can be seen only for T$\lesssim$70K and they are smeared for T$\gtrsim$150K.

The kink that emerges in the same range of $n_{2d}$ as the superconducting dome extends across a wide range of $V_{LG}$ ($3 \lesssim V_{LG} \lesssim 6$V) for $n_{2d}\gtrsim6\cdot10^{13}$cm$^{-2}$, and can be accessed only by LG biasing, due to the small capacitance of the solid BG. This prevents a continuous characterization of its behavior, as $n_{2d}$ induced by LG cannot be altered for $T\lesssim 220$K, as the ions are locked when the electrolyte is frozen. The kink that appears early in the metallic state, on the other hand, extends across a small range of $n_{2d}$ ($1\lesssim n_{2d} \lesssim 2\cdot10^{13}$cm$^{-2}$), and is ideally suited to be explored continuously by exploiting the dual-gate configuration.

We thus bias our samples in the low-density range of the metallic state ($n_{2d}\sim7\cdot 10^{12}$cm$^{-2}$) by applying V$_{LG}=0.9$V, and cool the system to 2.7K. We then apply $V_{BG}$ and fine-tune $n_{2d}$ across the kink. We constantly monitor $I_{BG}$ to avoid dielectric breakdown. Fig.\ref{figure:transport}c plots $\sigma_{2d}$ of a representative device subject to multiple $V_{BG}$ sweeps, as $n_{2d}$ is tuned across the kink. This reproduces well the behavior observed for low $V_{LG}$ ($1\lesssim n_{2d} \lesssim 2\cdot10^{13}$cm$^{-2}$). The hysteresis between increasing and decreasing $V_{BG}$ is minimal. This kink is suppressed by increasing T, similar to LG gating.

$V_{BG}$ provides us an independent tool to estimate $n_{2d}$: If $V_{LG}$ is small enough ($V_{LG}\lesssim 1$V) so that conduction in the channel can be switched off by sufficiently large negative $V_{BG}$ ($V_{BG}\lesssim -25$V), we can write $n_{2d}=C_{ox}/e \cdot (V_{BG}-V_{th})$. Here, $C_{ox}=\epsilon_{ox}/d_{ox}$ is the back gate oxide specific capacitance, $e=1.602\cdot10^{-19}$C is the elementary charge and $V_{th}$ is the threshold voltage required to observe a finite conductivity in the device. We neglect the quantum capacitance $C_q$ of MoS$_2$, since $C_q\gtrsim$100$\mu$F/cm$^2\gg C_{ox}$\cite{BrummePRB2015}. By using the dielectric constant of SiO$_2\,\epsilon_{ox}$=3.9\cite{ElKarehBook1995} and an oxide thickness $t_{ox}=90$nm (or $t_{ox}=285$nm, depending on the experiment) we obtain the $n_{2d}$ scale in the top axis of Fig.\ref{figure:transport}c, in good agreement with the corresponding values in Fig.\ref{figure:transport}a, estimated from the Hall effect measurements in Fig.\ref{figure:transport}b.

\begin{figure*}
\centerline{\includegraphics[width=0.8\textwidth]{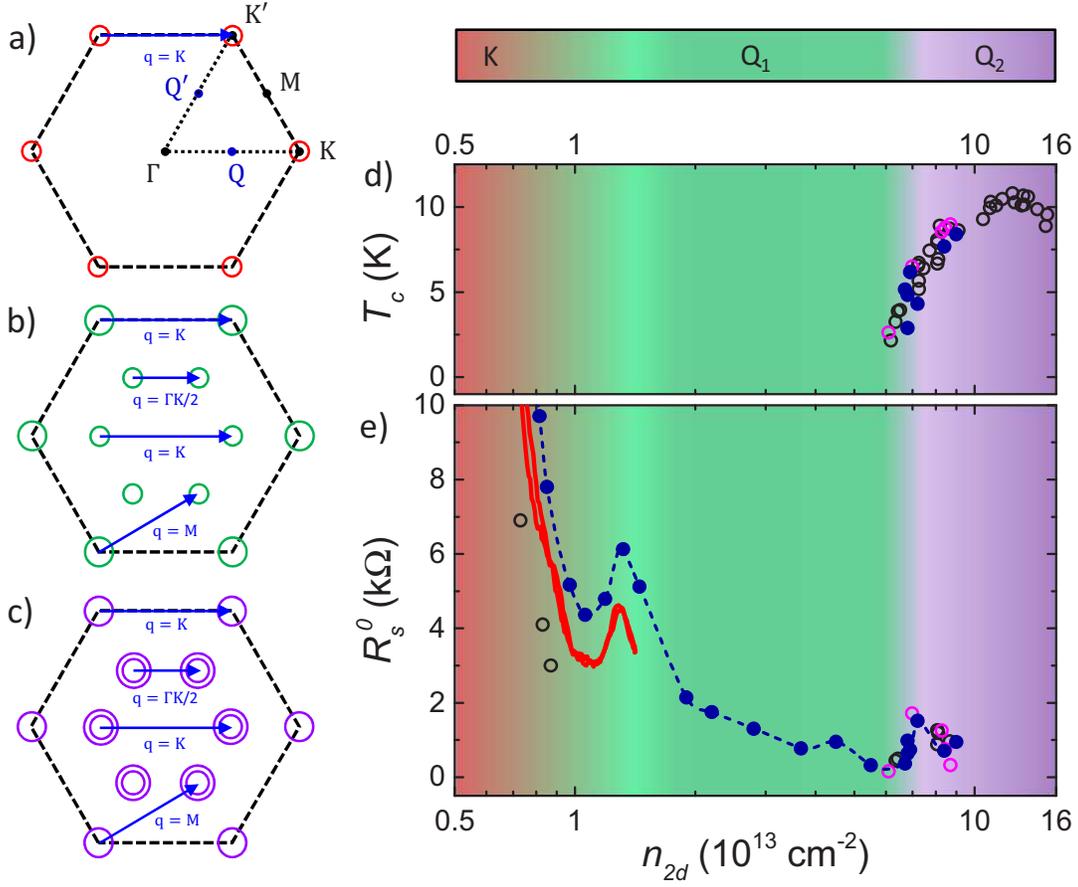}}
\caption{a-c) Fermi Surface of 3L-MoS$_2$ for the 3 doping values in Fig.\ref{figure:bands}b-d. High symmetry points $\Gamma$, M, K, K', and points Q and Q' are shown. Blue arrows indicate representative phonon wave vectors that connect the various FSs. d) SC dome of liquid-gated MoS$_2$ as a function of n$_{2d}$. $T_c$ is determined at $90\%$ of the total transition. e) $R_s^0$ as a function of $n_{2d}$, for increasing $V_{LG}$ (blue filled circles) and $V_{BG}$ (solid red line). In d,e), filled circles are our data, black and magenta open circles are taken from Refs.\citenum{YeScience2012, LuScience2015}. The background is color-coded to indicate the doping ranges highlighted in a-c).}
\label{figure:kinks}
\end{figure*}

The bandstructure of field-effect doped NL-MoS$_2$ depends on N\cite{BrummePRB2015} and strain\cite{BrummePRB2016}. A fully relaxed N-layer flake, with N$\geq$3, has been predicted to behave as follows\cite{BrummePRB2015, BrummePRB2016}: For small doping ($x\lesssim 0.05$e$^-$/unit cell, Figs.\ref{figure:bands}b and \ref{figure:kinks}a) only the two spin-orbit split sub-bands at K/K' are populated. At intermediate doping ($0.05\lesssim x \lesssim 0.1$ e$^-$/unit cell, Figs.\ref{figure:bands}c and \ref{figure:kinks}b), $E_F$ crosses the first spin-orbit split sub-band at Q/Q' (labeled Q$_1$). For large doping ($x\gtrsim 0.1$ e$^-$/unit cell, Figs.\ref{figure:bands}d and \ref{figure:kinks}c) $E_F$ crosses the second sub-band (Q$_2$) and both valleys become highly populated\cite{BrummePRB2015}. Even larger doping ($x\gtrsim 0.35$ e$^-$/unit cell) eventually shifts the K/K' valleys above $E_F$\cite{BrummePRB2015}.

When $E_F$ crosses these high-energy sub-bands at Q/Q', sharp kinks are expected to appear in the transconductance of gated FL-MoS$_2$\cite{BrummePRB2016} (see Fig.\ref{figure:bands}e). These are reminiscent of a similar behavior in liquid-gated FL graphene, where their appearance was linked to the opening of interband scattering channels upon the crossing of high-energy sub-bands\cite{YePNAS2011, Gonnelli2dMater2017, PiattiAppSS2017}. Even in the absence of energy-dependent scattering, Ref.\citenum{BrummePRB2016} showed that $\sigma_{2d}$ can be expressed as:
\begin{equation}
\sigma_{2d}=e^{2}\tau\langle v_{\parallel}^{2}\rangle N\left(E_{F}\right)\propto e^{2}\langle v_{\parallel}^{2}\rangle\label{eq:sigma_velocity}
\end{equation}
where $\tau\propto N(E_F)^{-1}$ is the average scattering time, and $N\left(E_{F}\right)$ is the density of states (DOS) at $E_F$. This implies that $\sigma_{2d}$ is proportional to the average of the squared in-plane velocity $\text{\ensuremath{\langle}}\ensuremath{v_{\parallel}^{2}}\text{\ensuremath{\rangle}}$ over the FS\cite{BrummePRB2016}. Since $\text{\ensuremath{\langle}}\ensuremath{v_{\parallel}^{2}}\text{\ensuremath{\rangle}}$ linearly increases with $n_{2d}$ and drops sharply as soon as a new band starts to get doped\cite{BrummePRB2016}, the kinks in $\sigma_{2d}$ (or, equivalently, $R_s$) at $T\lesssim15$K can be used to determine the onset of doping of the sub-bands in the Q/Q' valleys. At T=0, the kink is a sharp drop in $\sigma_{2d}$, emerging for the doping value at which $E_F$ crosses the bottom of the next sub-band. This correspondence is lost due to thermal broadening for T$>$0, leading to a smoother variation in $\sigma_{2d}$. If T is sufficiently large the broadening smears out any signature of the kinks, Fig.\ref{figure:transport}. Ref.\citenum{BrummePRB2016} calculated that, at finite T, the conductivity kinks define a \emph{doping range} where the sub-band crossing occurs (between $R_s$ minimum and maximum, i.e. the \emph{lower} and \emph{upper} bounds of each kink sets the resolution of this approach). Each sub-band crossing starts after the $R_s$ minimum at lower doping, then develops in correspondence of the inflection point, and is complete once the $R_s$ maximum is reached.

We show evidence for this behavior in Fig.\ref{figure:kinks}, where we plot $T_c$ (panel d) and $R_s^0$ (panel e) as a function of $n_{2d}$. The electric field is applied both in liquid-top-gate (filled dots and dashed line) and dual-gate (solid red line) configurations. For comparable values of $n_{2d}$, the liquid-gate geometry features larger $R_s^0$ than back-gated. This difference is due to increased disorder introduced when $n_{2d}$ is modulated via ionic gating\cite{Gonnelli2dMater2017, PiattiAppSS2017, PiattiAPL2017, GallagherNatCommun2015, OvchinnikovNatCommun2016}. Two kinks appear in the $n_{2d}$ dependence of $R_s$: a low-doping one for $1.5\cdot 10^{13} \lesssim n_{2d} \lesssim 2\cdot 10^{13}$cm$^{-2}$, and a high-doping one for $7\cdot 10^{13} \lesssim n_{2d} \lesssim 9\cdot 10^{13}$cm$^{-2}$. The plot of the SC dome of gated MoS$_2$ on the same $n_{2d}$ scale  shows that the low-doping kink appears well before the SC onset, while the second appears immediately after, before the maximum $T_c$ is reached.

These results can be interpreted as follows. When $n_{2d} \lesssim 1\cdot10^{13}$cm$^{-2}$, only the spin-orbit split sub-bands at K/K' are populated, and the FS is composed only by two pockets, Fig.\ref{figure:kinks}a. For $n_{2d}$ between$\sim1.5$ and $2\cdot 10^{13}$cm$^{-2}$, $E_F$ crosses the bottom of the Q$_1$ sub-band and two extra pockets appear in the FS at Q/Q'\cite{GePRB2013,BrummePRB2015}, Fig.\ref{figure:kinks}b. The emergence of these pockets induces a Lifshitz transition, i.e. an abrupt change in the topology of the FS\cite{LifshitzJETP1960}. Once Q$_1$ is populated and $E_F$ is large enough ($n_{2d}\sim6\cdot 10^{13}$cm$^{-2}$), the system becomes superconducting\cite{LuScience2015, YeScience2012}. For slightly larger $E_F$ ($7\cdot 10^{13} \lesssim n_{2d} \lesssim 9\cdot 10^{13}$cm$^{-2}$), $E_F$ crosses the bottom of Q$_2$ resulting in a second Lifshitz transition, and other two pockets emerge in the FS at Q/Q'\cite{BrummePRB2015}, Fig.\ref{figure:kinks}c.

We note that the experimentally observed kinks are at different $n_{2d}$ with respect to the theoretical ones for 3L-MoS$_2$\cite{BrummePRB2016}. Ref.\citenum{BrummePRB2016} predicted that for a $1.28\%$ in-plane tensile strain, Q$_1$ and Q$_2$ should be crossed for $n_{2d}\sim5\cdot 10^{13}$ and $\sim1\cdot 10^{14}$. Since the positions of the sub-band crossings are strongly dependent on strain\cite{BrummePRB2016}, we estimate the strain in our devices by monitoring the frequency of the E$_{2g}^1$ mode via Raman spectroscopy.
\begin{figure}
\centerline{\includegraphics[width=\columnwidth]{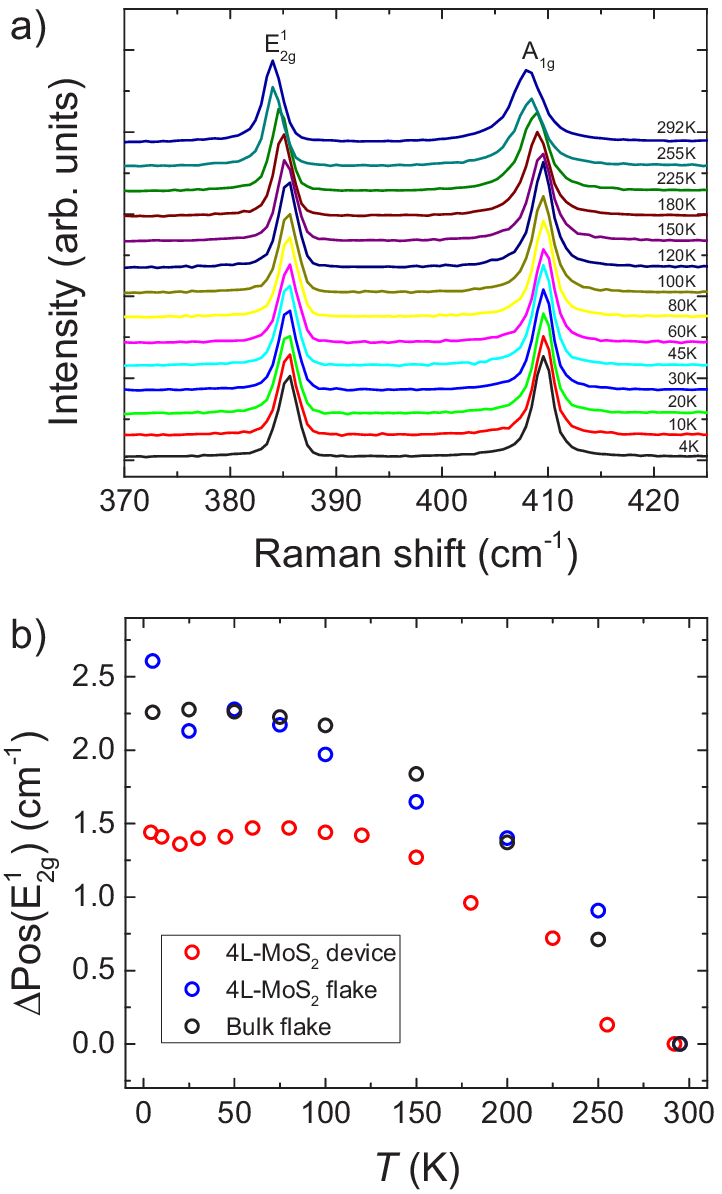}}
\caption{a) Raman spectra of the 4L-MoS$_2$ device in Fig.\ref{figure:device}c from 4 to 292K. b) Shift in the position of the E$_{2g}^1$ mode as a function of T for as-prepared bulk flake (black circles), a 4L-MoS$_2$ flake (blue circles), and a 4L-MoS$_2$ device with Au contacts (red circles).}
\label{fig:strain}
\end{figure}
Strain can arise due to a mismatch in the thermal expansion coefficients (TECs) of MoS$_2$\cite{GanPRB2016}, SiO$_2$ substrate\cite{NIST_standard1991} and Au electrodes\cite{NixPR1941}. Upon cooling, MoS$_2$, SiO$_2$ and Au would normally undergo a contraction. However the flake is also subject to a tensile strain due to TEC mismatch\cite{YoonNL2011}. The strain, $\epsilon_{MoS_2}$, due to the MoS$_2$-SiO$_2$ TEC mismatch is:
\begin{equation}
\epsilon_{MoS_2}=\int_{T}^{292K} [\alpha_{MoS_2}(T)-\alpha_{SiO_2}(T)] dT
\end{equation}
whereas the strain, $\epsilon_{Au}$, due to the Au contacts is:
\begin{equation}
\epsilon_{Au}=\int_{T}^{292K} [\alpha_{Au}(T)-\alpha_{SiO_2}(T)] dT
\end{equation}
$\epsilon_{MoS_2}$ and $\epsilon_{Au}$ are$\sim$0.1\% and$\sim$0.3\% at $\sim$4K, respectively\cite{YoonNL2011}.

Any FL-MoS$_2$ on SiO$_2$ will be subject to $\epsilon_{MoS_2}$ at low T. When the flake is contacted, an additional contribution is present due to $\epsilon_{Au}$. This can be more reliably estimated performing T-dependent Raman scattering and comparing the spectra for contacted and un-contacted flakes\cite{YoonNL2011,MohiPRB2009}. Figs.\ref{fig:strain}a,b show how a T decrease results in the E$_{2g}^1$ mode shifting to higher frequencies for both as-prepared and contacted 4L-MoS$_2$, due to anharmonicity\cite{KlemPR1966}. However, in the as-prepared 4L-MoS$_2$, the up-shift is$\sim$1cm$^{-1}$ larger with respect to the contacted one. This difference points to a further tensile strain. Refs.\citenum{LeeNC2017, ConlNL2013} suggested that uniaxial tensile strain on 1L-MoS$_2$ induces a E$_{2g}^1$ softening and a splitting in two components: E$_{2g}^{1+}$ and E$_{2g}^{1-}$\cite{LeeNC2017, ConlNL2013}. The shift rates for E$_{2g}^{1+}$ and E$_{2g}^{1-}$ are from -0.9 to -1.0cm$^{-1}$/\% and from -4.0 to -4.5cm$^{-1}$/\%, respectively\cite{LeeNC2017, ConlNL2013}. We do not observe splitting, pointing towards a biaxial strain. As for Ref.\citenum{MohiPRB2009}, we calculate a shift rate of E$_{2g}^1$ for biaxial strain from -7.2 to -8.2cm$^{-1}$/\%. The amount of tensile strain on the 4L-MoS$_2$ device can thus be estimated. The E$_{2g}^1$ up-shift difference between contacted and as-prepared 4L-MoS$_2$, $\Delta$Pos(E$_{2g}^1$), at 4K is$\sim$-1.0cm$^{-1}$, corresponding to an additional $\sim$0.13\% biaxial tensile strain. Thus, assuming a 0.1\% strain for the as-prepared 4L-MoS$_2$ due to TEC mismatch with SiO$_2$, we estimate the total strain in the contacted 4L-MoS$_2$ to be$\sim$0.23\% at$\sim$4K.
\begin{figure}
\centerline{\includegraphics[width=\columnwidth]{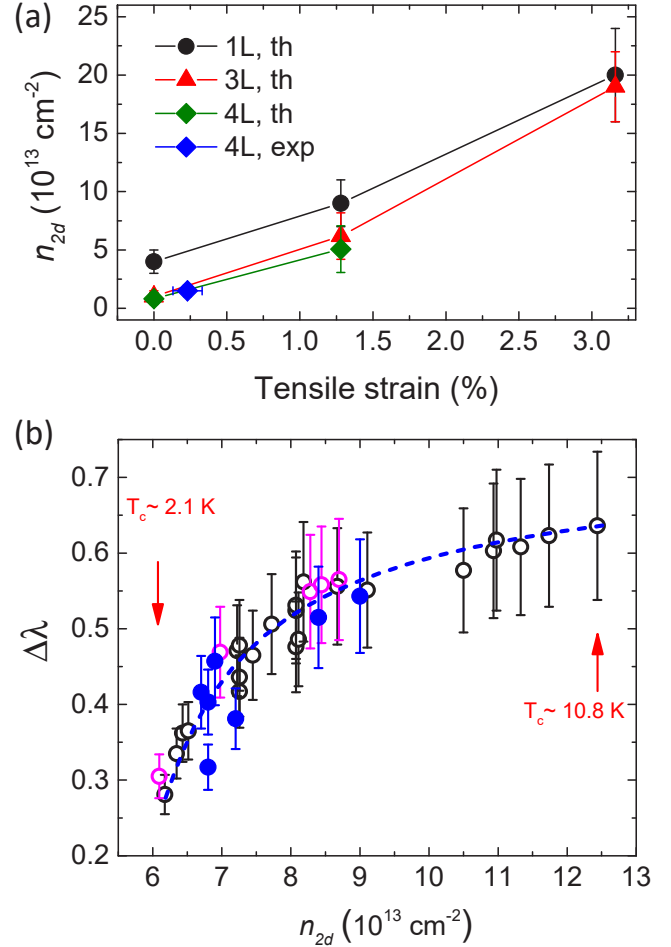}}
\caption{a) Surface carrier densities required to cross the Q$_1$ sub-band in FL-MoS$_2$ as a function of tensile strain. Theoretical values for 1L (black dots and line) and 3L (red triangle and line) from Ref.\citenum{BrummePRB2016}; values for 4L (green diamonds and line) are by linear extrapolation. Blue diamond is the present experiment. b) EPC enhancement due to the crossing of the Q$_2$ sub-band, $\Delta \lambda$, as a function of $n_{2d}$, assuming $\omega_{ln}=230\pm30$K and $\mu^*=0.13$\cite{GePRB2013}. Filled blue circles are our experiments. Black and magenta open circles from Refs.\citenum{YeScience2012, LuScience2015}. The blue dashed line is a guide to the eye.}
\label{figure:coupling}
\end{figure}

Fig.\ref{figure:coupling}a shows that, for $0.23$\% tensile strain, the experimentally observed positions of the kinks agree well with a linear extrapolation of the data of Ref.\citenum{BrummePRB2016} to 4L-MoS$_2$ (representative of our experiments) and for in-plane strain between $0\%$ (bulk) and $1.28\%$ (fully relaxed). These findings indicate that, while the mechanism proposed in Ref.\citenum{GePRB2013} qualitatively describes the general behavior of gated FL-MoS$_2$, quantitative differences arise due to the spin-orbit split of the Q$_1$ and Q$_2$ sub-bands. The main reason for the EPC (and, hence, $T_c$) increase is the same, i.e. the increase in the number of phonon branches involved in the coupling when the high-energy valleys are populated\cite{GePRB2013}. However, the finite spin-orbit-split between the sub-bands significantly alters the FS connectivity upon increasing doping\cite{BrummePRB2015}. If we consider the relevant phonon wave vectors ($q$=$\Gamma$,K,M,$\Gamma$K$/2$) for 1L- and FL-MoS$_2$\cite{MolinaPRB2011,AtacaJPCC2011}, and only the K/K' valleys populated, then only phonons near $\Gamma$ and K can contribute to EPC\cite{GePRB2013}. The former strongly couple e$^-$ within the same valley\cite{GePRB2013}, but cannot contribute significantly due to the limited size of the Fermi sheets\cite{GePRB2013}. The latter couple e$^-$ across different valleys\cite{GePRB2013}, and provide a larger contribution\cite{GePRB2013}, insufficient to induce SC. MoS$_2$ flakes are metallic but not superconducting before the crossing of Q$_1$. When this crossing happens, the total EPC increases due to the contribution of longitudinal phonon modes near K\cite{GePRB2013} (coupling states near two different Q or Q'), near $\Gamma$K$/2$\cite{GePRB2013} (coupling states near Q to states near Q'), and near M\cite{GePRB2013} (coupling states near Q or Q' to states near K or K'). However, this first EPC increase associated with Q$_1$ is not sufficient to induce SC, as the SC transition is not observed until immediately before the crossing of the spin-orbit-split sub-band Q$_2$ and the second doping-induced Lifshitz transition. Additionally, the SC dome shows a maximum in the increase of $T_c$ with doping ($dT_c/dn_{2d}$) across the Q$_2$ crossing, i.e. when a new FS emerges. Consistently, the subsequent reduction of $T_c$ for $n_{2d}\geq 13\cdot 10^{13}$cm$^{-2}$ can be associated with the FS shrinkage and disappearance at K/K'\cite{GePRB2013, BrummePRB2015}, and might also be promoted by the formation of an incipient Charge Density Wave\cite{RosnerPRB2014, Piattiarxiv2018} (characterized by periodic modulations of the carrier density coupled to a distortion of the lattice structure\cite{GrunerBook2009}).

Since the evolution of the bandstructure with doping is similar in several semiconducting TMDs\cite{BrummePRB2015, DasPRB2015, BrummePRB2016, WuNatCommun2016, KangNanoLett2017}, this mechanism is likely not restricted to gated MoS$_2$. The $T_c$ increase in correspondence to a Lifshitz transition is reminiscent of a similar behavior observed in CaFe$_2$As$_2$ under pressure\cite{GonnelliSciRep2016}, suggesting this may be a general feature across different classes of materials.

We note that the maximum $T_c\sim11$K is reached at $n_{2d}\simeq 12\cdot 10^{13}$cm$^{-2}$, as reported in Ref.\citenum{YeScience2012}. This is a doping level larger than any doping level which can be associated with the kink. Thus, the Q$_2$ sub-band must be highly populated when the maximum $T_c$ is observed. We address this quantitatively with the Allen-Dynes formula\cite{AllenDynes}, which describes the dependence of $T_c$ by a numerical approximation of the Eliashberg theory accurate for materials with a total $\lambda\lesssim1.5$\cite{AllenDynes}:
\begin{equation}
\label{EqAllenDynes}
T_c(n_{2d})=\frac{\omega_{ln}}{1.2}\mathrm{exp}\left\lbrace\frac{-1.04\left[1+\lambda(n_{2d})\right]}{\lambda(n_{2d})-\mu^*\left[1+0.62\lambda(n_{2d})\right]}\right\rbrace
\end{equation}
where $\lambda(n_{2d})$ is the total EPC as a function of doping, $\omega_{ln}$ is the representative phonon frequency and $\mu^*$ is the Coulomb pseudo-potential. It is important to evaluate the increase in EPC between the non-superconducting region ($n_{2d}\lesssim6\times10^{13}$cm$^{-2}$) and the superconducting one, i.e. the enhancement in $\lambda$ due to the crossing of the sub-band at Q$_2$. $\Delta\lambda=\lambda(T_c)-\lambda(T_c=0)$ indicates the EPC increase due to the appearance of e$^-$ pockets at Q$_2$. By setting $\omega_{ln}=230\pm30$K and $\mu^*=0.13$ (as for Ref.\citenum{GePRB2013}), and using Eq.\ref{EqAllenDynes}, we find that the limit of $\lambda(T_c)$ for $T_c\to0$ is $\sim0.25$. The corresponding $\Delta\lambda$ vs. $n_{2d}$ dependence is shown in Fig.\ref{figure:coupling}b. The crossing at Q$_2$ results in a maximum $\Delta\lambda=0.63\pm0.1$, with a maximum EPC enhancement of $350\pm40\%$ with respect to the non-superconducting region. This indicates that the largest contribution to the total EPC, hence to the maximum $T_c\sim11$K, is associated with the population of the Q$_2$ sub-band. This is consistent with the reports of a reduced $T_c\sim2$K in 1L-MoS$_2$\cite{CostanzoNatNano2016, FuQuantMater2017}, shown to be superconducting for smaller $n_{2d}\sim5.5\cdot10^{13}$cm$^{-2}$\cite{FuQuantMater2017}, hence likely to populate Q$_1$ only. $n_{2d}\sim5\cdot10^{13}$cm$^{-2}$ is also the doping expected for the crossing of Q$_1$ in 1L-MoS$_2$ in presence of a low-T strain similar to that in our 4L-MoS$_2$ devices (see Fig.\ref{figure:coupling}a).

In summary, we exploited the large carrier density modulation provided by ionic gating to explore sub-band population and multivalley transport in MoS$_2$ layers. We detected two kinks in the conductivity, associated with the doping-induced crossing of the two sub-bands at Q/Q'. By comparing the emergence of these kinks with the doping dependence of $T_c$, we showed how superconductivity emerges in gated MoS$_2$ when the Q/Q' valleys are populated, while previous works only considered the filling of K/K'. We highlighted the critical role of the population of the second spin-orbit-split sub-band, Q$_2$, (and the consequent increase of the FS available for EPC) in the appearance of superconductivity and in the large enhancement of $T_c$ and of EPC in the first half of the superconducting dome. Our findings explain the doping dependence of the SC state at the surface of gated FL-MoS$_2$, and provide a key insight for other semiconducting transition metal dichalcogenides.
\bigskip
\subsection*{Acknowledgments}
We thank M. Calandra for useful discussions. We acknowledge funding from EU Graphene Flagship, ERC Grant Hetero2D, EPSRC Grant Nos. EP/509K01711X/1, EP/K017144/1, EP/N010345/1, EP/M507799/ 5101, and EP/L016087/1 and the Joint Project for the Internationalization of Research 2015 by Politecnico di Torino and Compagnia di San Paolo.

The authors declare no competing financial interests.

\end{document}